\documentclass[prd,showkeys,floatfix,nofootinbib,
               preprint,12pt,
               tightenlines,fleqn]{revtex4-1} %for preprint

\usepackage{amsmath,amssymb,revsymb,graphicx,dcolumn}
\usepackage{leftidx}

\usepackage{marginnote}

\newcommand{\beq}{\begin{equation}}
\newcommand{\eeq}{\end{equation}}
\newcommand{\beqa}{\begin{eqnarray}}
\newcommand{\eeqa}{\end{eqnarray}}
\newcommand{\bsubeqs}{\begin{subequations}}
\newcommand{\esubeqs}{\end{subequations}}

\usepackage{mwe}
\newcommand{\imineq}[2]{\vcenter{\hbox{\includegraphics[height=#2ex]{#1}}}}

\begin{document}
\title[]
      {On free fall of fermions and antifermions}
\author{Viacheslav A. Emelyanov}
\email{viacheslav.emelyanov@partner.kit.edu}
\affiliation{Institute for Theoretical Physics,\\
Karlsruhe Institute of Technology,\\
76131 Karlsruhe, Germany\\}

\begin{abstract}
\vspace*{2.5mm}\noindent
We propose a model describing spin-half quantum particles in curved spacetime in the framework~of
quantum field theory.~Our~model~is~based on embodying Einstein's equivalence principle and general 
covariance in the definition of quantum-particle states.~With this model at hand,~we compute~several
observables which characterise spin-half quantum particles in a gravitational field.~In particular,~we find
that spin precesses in a normal Fermi frame,~even in the absence of torsion.~The~effect~appears~to
be complementary to free-fall non-universality we have recently reported about for spinless quantum
particles.~Furthermore,\,we find that quantum-particle gravitational-potential~energy~is~insensitive~to
wave-packet spreading in the Earth's gravitational field,~that is responsible~for~the~non-universality
of free fall in quantum theory.~This theoretical result provides another channel for the experimental
study of our quantum-particle model by using gravitational spectrometers.~Finally,~we also find~that
(elementary) fermions and antifermions are indistinguishable in gravity.
\end{abstract} 

\keywords{}

\maketitle

\section{Introduction}
\label{sec:introduction}

There are several research fields in modern theoretical physics, focusing on various~aspects
of quantum field theory.~This primarily finds its application in elementary particle physics.~In
fact,~quantum theory of fields was originally developed to model electromagnetic~interaction
to naturally require a unification of quantum mechanics and~special~relativity.~Nowadays,~the
Standard Model of elementary particle physics successfully involves the quantum-field-theory
formalism for modelling high-energy scattering processes taking place in colliders.

In the framework of general relativity,~however,~special relativity is a special-case~theory.~It
is in conflict with observations whenever gravity cannot be neglected with respect to the~rest
three fundamental interactions.~In particular,~the free-fall observation~of
neutrons~\cite{McReynolds,Dabbs&Harvey&Paya&Horstmann,Koester}~shows
that the Standard Model has to be extended to comprise general relativity.~This~circumstance
leads to a problem of field quantisation in curved spacetime.

This is problematic because it is not self-evident how to properly generalise quantum field
theory over Minkowski spacetime to a non-Minkowski one.~The basic reason of that~is~the~role
which the Poincar\'{e} group plays in the definition of quantum vacuum and, correspondingly, of
Fock space in field quantisation in Minkowski
spacetime~\cite{Weinberg,Peskin&Schroeder,Itzykson&Zuber,Bogolyubov&Shirkov,Srednicki}.~The
basic idea,~which has been put forward in this regard, is to utilise isometry group
of a given non-Minkowski spacetime for that
purpose~\cite{DeWitt,Birrell&Davies,Parker&Toms}.~This particularly implies global field
quantisation,~because~this~approach
demands the knowledge of metric tensor at all space-time points.~Apart from this is unfeasible
in practice, it is unknown if the observable Universe has to have any particular exact~isometry
group.~Thus, global field quantisation assumes the usage of space-time geometries which have
a non-trivial isometry group.~These geometries must approximate~some~space-time~regions~of
the observable Universe.~In other words,~their isometry groups can be thought of as local~non-
exact symmetries of the Universe.~The Einstein equivalence principle states that the~Poincar\'{e}
group can be considered as its local non-exact symmetry too~\cite{Will}.~The question~then~arises~as
to whether global field quantisation locally reduces to that used in particle physics.

The Standard Model of particle physics has been in fact being tested in the presence of the
Earth's gravitational field.~It may be approximately described by Schwarzschild~spacetime if
the Earth's rotation is neglected.~The Schwarzschild metric is invariant with respect~to time
translations and rotations.~Although these form a subgroup of the Poincar\'{e} group,~there~is~no
guarantee that global field quantisation based on the isometry group of Schwarzschild~spacetime
locally agrees with global field quantisation in Minkowski spacetime.~This is because~the
Schwarzschild-time and -space coordinates differ from local Minkowski-time and -space ones.
Indeed,~general coordinates $x$ and normal Riemann coordinates $y$ in the neighbourhood~of~$X$
are related as follows~\cite{Petrov}:
\beqa\label{eq:local-inertial-coordinates}
y^c(x) &\approx& Y^c + (x-X)^c + \frac{1}{2}\,\Gamma_{ab}^c\,(x-X)^a(x-X)^b\,,
\eeqa
where the Latin indices run over $\{0,1,2,3\}$,
$\Gamma_{ab}^c$ are Christoffel symbols computed at $X$ and higher-order
terms with respect~to metric derivatives have been neglected on the right-hand side
of~\eqref{eq:local-inertial-coordinates}.~So,~Schwarzschild spacetime is locally spherical symmetric
at any $X$,~which~is~due to Einstein's equivalence principle, while globally spherical symmetric
with respect to a single point which is known as the central singularity of the Schwarzschild
geometry.~Furthermore,
the Schwarzschild time does asymptotically match a local Minkowski time
at spatial~infinity. However,~this is non-existent in practice.~That is,~the Schwarzschild-time
translations~do~not correspond to local-Minkowski-time translations if metric derivatives
are non-zero.

The successful application of quantum field theory in elementary particle physics instructs us
to consider~plane-wave modes (used to expand a quantum
field~\cite{Weinberg,Peskin&Schroeder,Itzykson&Zuber,Bogolyubov&Shirkov,Srednicki}), i.e.
\beqa\label{eq:local-plane-wave}
\psi_{P}(y) &=& \exp\big({-}iP_a\,y^a/\hbar\big)\,,
\eeqa
in the Riemann inertial frame,~where $P$ is the on-mass-shell $4$-momentum:
$\eta_{ab}P^aP^b = (Mc)^2$, where $c$ is the speed of light in vacuum in the absence of
external fields~\cite{Weinberg,Peskin&Schroeder,Itzykson&Zuber}.~We~then~obtain
from~\eqref{eq:local-inertial-coordinates} for $\mathbf{X}$ being at the Earth's surface~that
\beqa\label{eq:local-plane-wave-phase}
\psi_{P}(y)\big|_{|\mathbf{P}| \,\ll\, Mc} &\propto& \exp\big({-}iMc^2(1 + g_\oplus z/c^2)t/\hbar\big)\,,
\eeqa
where $g_\oplus \approx 9.81\,\text{m}/\text{s}^2$ is the free-fall acceleration, $z \equiv (x{-}X)^3$
is the vertical height above~the Earth's surface and $t \equiv (x{-}X)^0/c$ is the Schwarzschild time.~The
$g_\oplus$-dependent~term~in~\eqref{eq:local-plane-wave-phase}
has been derived in~\cite{Colella&Overhauser} from the Schr\"{o}dinger equation with Newton's
gravitational potential. This term gives rise to gravity-induced quantum interference which has
been observed in~the Colella--Overhauser--Werner
experiment~\cite{Colella&Overhauser&Werner}.~It is a non-inertial-frame effect,~that~also~shows
up in an accelerated frame, as that has been confirmed by the Bonse--Wroblewski experiment
\cite{Bonse&Wroblewski}.~These experiments demonstrate that quantum interference cannot be used
to distinguish between uniform gravity and acceleration, in accord with Einstein's principle~\cite{Nauenberg}.

In the Schwarzschild frame, global field quantisation instructs us to consider modes which
are eigenfunctions of the Killing vector generating the Schwarzschild-time
translations~\cite{DeWitt,Birrell&Davies,Parker&Toms}:
\beqa
\psi_{P}(x)\big|_{|\mathbf{P}| \,\ll\, Mc} &\propto& \exp\big({-}iMc^2t/\hbar\big)\,,
\eeqa
where the coefficient of proportionality is independent of the Schwarzschild-time coordinate
for any value of $|\mathbf{P}|$.~The mode phase differs from that
of~\eqref{eq:local-plane-wave-phase}.~So,~the physical meaning~of~Fock space based on these
modes is obscure in light of the experiments mentioned above.

From other side,~the plane-wave modes~\eqref{eq:local-plane-wave} cannot be exact solutions of
a field equation~in curved spacetime.~Still,~these can be treated as approximate solutions in local
inertial~frames. These frames can generically be introduced~relative~to either
a point or a trajectory.~However, a quantum-field operator,~e.g.~$\hat{\Psi}(x)$,~depends on a single
point.~In other~words,~$\hat{\Psi}(x)$~is~a~local operator, whereas the 
coordinate transformation $x \rightarrow y(x)$ means that $y$ depends~on~a~pair~of points
-- $x$ and $X$.~This might in turn mean that one
has to~consider~$\hat{\Psi}(x) \rightarrow \hat{\Psi}_X(y)$~under~the coordinate
transformation\,\eqref{eq:local-inertial-coordinates}.\,Theoretical particle physics
relies~on~the~Minkowski-spacetime approximation,~where one deals with $\hat{\Psi}(y)$
which does not depend on 
$X$~\cite{Weinberg,Peskin&Schroeder,Itzykson&Zuber,Bogolyubov&Shirkov,Srednicki}.~These~all~might
mean that $\hat{\Psi}(x)$ itself should appear in
observables~in such a way that~$X$~does~not show up~if the space-time curvature is
neglected.~It~is~in~fact required for Einstein's equivalence principle to hold in local quantum phenomena.

One of these phenomena is scattering of particles.~At tree level of perturbation theory,~the
probability amplitude covariantly generalised to curved spacetime pictorially~reads~for~a~pair
of charged particles in the framework of quantum electrodynamics as follows:
\beqa\label{eq:probability-amplitude}
\mathbf{\imineq{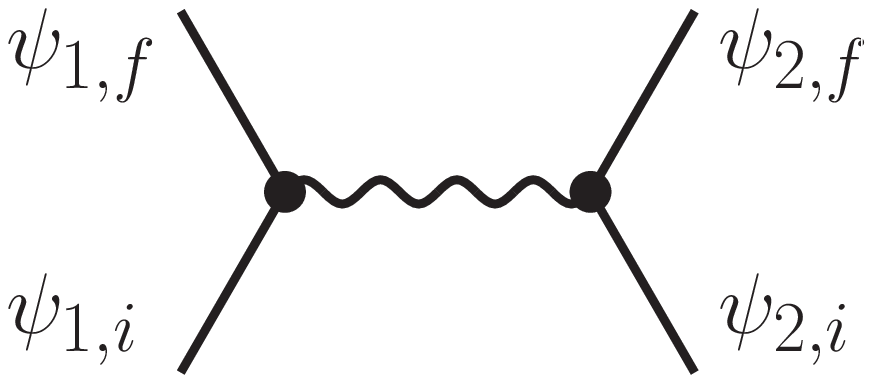}{9.5}}\hspace{-1mm} &\propto& {\prod\limits_{s \,\in\, \{1,2\}}}
{\int}d^4x_s\sqrt{-g(x_s)}\,
J_1^\mu(x_1)\,G_{\mu\nu}^F(x_1,x_2)\,J_2^\nu(x_2)\,,
\eeqa
where $G_{\mu\nu}^F(x_1,x_2)$ is a Feynman propagator of the photon field in curved
spacetime,~naturally depending on the geodesic distance between $x_1$ and
$x_2$ and the space-time curvature~\cite{Birrell&Davies},~and
\beqa\label{eq:current-density}
J_s^{\mu}(x) &\equiv&
\langle \psi_{s,f}|\hat{J}_s^\mu(x)|\psi_{s,i}\rangle
- \langle\psi_{s,f}|\psi_{s,i}\rangle\langle\Omega|\hat{J}_s^\mu(x)|\Omega\rangle\,,
\eeqa
where $\hat{J}_s^\mu(x)$ is the current-density operator~\cite{Peskin&Schroeder}
and $|\Omega\rangle$ stands for the quantum-vacuum~state.
This probability amplitude is diffeomorphism invariant if and only if the particle
states $|\psi_{s,i}\rangle$ and $|\psi_{s,f}\rangle$ are independent of coordinate
frames.~Besides,~$\langle\psi_{s,f}|\psi_{s,i}\rangle$ turns~identically~to~zero
if the initial and final states of a particle are orthogonal to each other.~In practice,~this~might
not be the case for free particles if one deals with superpositions of plane waves in Minkowski
spacetime.~Alternatively,~one may replace $\hat{J}_s^\mu(x)$ by ${:}\hat{J}_s^\mu(x){:}$ for which
$\langle\Omega|{:}\hat{J}_s^\mu(x){:}|\Omega\rangle =0$~\cite{Weinberg,Peskin&Schroeder,Itzykson&Zuber,Bogolyubov&Shirkov,Srednicki}.
In any case,~the current density $J_s^{\mu}(x)$ defined in~\eqref{eq:current-density} is localised in
a spacetime~region~in~which the states $|\psi_{s,i}\rangle$ and $|\psi_{s,f}\rangle$ overlap.~It means
that the probability amplitude~\eqref{eq:probability-amplitude} locally~reduces to that computed
by using the Minkowski-spacetime approximation if the in-coming particles
are brought to a space-time region of size being much smaller than a local curvature~length~in
that region.~If otherwise,~the probability amplitude tends to zero by increasing 
the distance between the localisation regions of $J_1^\mu(x)$ and $J_2^\mu(x)$,~following from
the cluster decomposition principle (see Ch.~4 in~\cite{Weinberg} and~\cite{Emelyanov&Klinkhamer}
for concrete computations).

In theoretical particle physics, one commonly deals with in-coming states depending~solely on
initial momenta.~In classical theory,~it is additionally required to set their initial~positions
in order to ascertain particles' trajectories.~This has to be the case in~quantum theory~as~well.
Clearly,~this does not contradict to Heisenberg's uncertainty~principle.~It means~that~we~need
to replace $|\psi_{i}\rangle$ by $|\psi_{X,P}\rangle$,~where $|\psi_{X,P}\rangle$ propagates over
a trajectory passing through~$(X,P)$~in the phase space.~In Minkowski spacetime,~this
trajectory must be $x(\tau) = X + (P/M)\,\tau$, where $M$ is particle's mass and $\tau$ denotes
proper time.~Therefore, in the coordinate representation, $|\psi_{X,P}\rangle$ turns into the
wave function $\psi_{X,P}(x)$ to depend on $x$ and $X$.~Such wave functions~enter
the current densities~\eqref{eq:current-density}.~This line of reasoning shows,~thereby,~that all
quantities entering~the scattering amplitude depend not on absolute positions, 
like quantum-field operators, $\hat{\Psi}_s(x)$, but rather on relative positions.

To summarise,~quantum-field operators are fundamental objects in quantum~field~theory.
Still,~quantum phenomena manifest themselves through the interaction of quantum~particles.
A quantum-particle model based on modes $\psi_P(x)$ is not complete,~because~initial~position~$X$
has to be part of the model,~otherwise this is at odds with Bohr's correspondence~principle.~It
means that the quantum-field expansion over~modes $\psi_P(x)$ should play no underlying role~for
the description of quantum particles (cf.~\cite{DeWitt,Birrell&Davies,Parker&Toms}).~Instead of
that,~one should look for $\psi_{X,P}(x)$ to provide the wave-function description for quantum
particles in the weak-gravity limit.~This naturally allows us to fulfil the condition that 
$\psi_{X,P}(x)$ reduces to a plane-wave superposition in a local
Minkowski frame for $x$ being close to $X$~\cite{Emelyanov-2020,Emelyanov-2021,Emelyanov-2022},~in
accordance with the application of quantum field theory in particle
physics~\cite{Weinberg,Peskin&Schroeder,Itzykson&Zuber,Bogolyubov&Shirkov,Srednicki}.~For this
condition to be fulfilled in~any~local Minkowski frame, $\psi_{X,P}(x)$ should,~additionally,~transform
as a zero-rank tensor under general coordinate transformations.~This explains
how quantum field theory~based~on~the~Minkowski-
spacetime approximation and high-energy experiments
done in the Earth's gravitational field can locally reconcile with each other.

We intend here to study spin-half quantum particles in curved spacetime
in~the~framework of quantum~field theory.~Our idea is here
to build Einstein's equivalence principle~and~general covariance~\cite{Einstein}~into~the~model
of quantum particles in gravity,~that~we~have~lately~put~forward
in~\cite{Emelyanov-2020,Emelyanov-2021,Emelyanov-2022}.\,Accordingly,~we~will~construct
and study a quantum spin-half~state which~is~locally 
represented by a superposition of positive-frequency plane waves,~as in Minkowski~spacetime,
and which is invariant under general coordinate transformations.\,In an arbitrary~curved~space- 
time, it is hard to construct such a state non-perturbatively in curvature.~For this reason,~we
will do this in perturbation theory by only taking the leading-order curvature contribution~to
the quantum spin-half~state into account.~This approximation proves to be sufficient to reach
the main~goal of the article,~that is to compare the kinematics of a quantum particle modelled
by such a state with that of a classical particle of same mass and spin.~This analysis~allows~us
to determine possible experimental setups which are suitable for testing our model.

Throughout, we use natural units $c = G = \hbar = 1$, unless otherwise stated.

\section{Covariant spin-half particle}

\subsection{Minkowski spacetime}

Here we want to model a spin-half quantum particle in Minkowski spacetime.~We~consider
that the particle is initially placed at $Y$ in position space and at $P$ in momentum space.~So,~it
is modelled by the following state:
\beqa
|\psi_{Y,P}\rangle &\equiv& \hat{a}^\dagger(\psi_{Y,P})|\Omega\rangle\,,
\eeqa
where $|\Omega\rangle$ is the quantum vacuum and $\hat{a}^\dagger(\psi_{Y,P})$
is an operator creating the spin-half quantum particle to be characterised by the
wave function $\psi_{Y,P}(y)$. 

First, theoretical particle physics defines a unique quantum vacuum which is known in~the
literature as the Minkowski vacuum~\cite{Weinberg,Peskin&Schroeder,Itzykson&Zuber,Bogolyubov&Shirkov,Srednicki}.~It
is unique with respect to the Poincar\'{e}~isometry
group of Minkowski spacetime.~However,~the observable Universe may have no~exact isometry
group.~Yet,~it does have approximate isometry groups emerging at different length scales.~One
of the physically relevant examples is a \emph{local} Poincar\'{e} group.~This emerges in~a~vicinity~of~any
non-singular space-time point at length scales much smaller than a local~curvature~length~at
that point.~This statement is a consequence of Einstein's equivalence principle which~is~so~far
in agreement with observations.~At cosmological length scales, the Universe looks as~de-Sitter
spacetime~\cite{Mukhanov}.~There is no unique quantum vacuum in de-Sitter spacetime,~which
is invariant under the de-Sitter isometry group~\cite{Birrell&Davies}.~In general, there is no preferred
procedure~to~choose~a unique quantum vacuum in curved spacetime.~In the absence of any exact
universe isometry group, $|\Omega\rangle$ is, at least, supposed to be unitarily equivalent to all local
Minkowski vacua~which can be defined in local inertial frames.~This is required for having a locally
unique Fock space in the observable Universe.~In other words,~radio waves coming from Sagittarius
$\text{A}^\star$~must~have the same nature like those produced here on the Earth.

Second, the particle creation operator is defined as follows:
\beqa\label{eq:particle-creation-minkowski}
\hat{a}^\dagger(\psi_{Y,P}) &\equiv&
{\int}d^3\mathbf{y}\,\hat{\Psi}^\dagger(y)\,\psi_{Y,P}(y)\,,
\eeqa
where $\hat{\Psi}(y)$ is a Dirac field satisfying Dirac's equation with the mass term $M$.~It is
common~in theoretical particle physics to consider a definite momentum wave function,~namely
\beqa\label{eq:non-normalisable-wave-function}
\big(\psi_{Y,P}(y)\big)_\textrm{non-normalisable} &=& u(P)\,e^{-iP{\cdot}(y-Y)}\,,
\eeqa
where $u(P)$ is a 4-dimensional column vector
(see Sec.~3.3 in~\cite{Peskin&Schroeder}).~Nevertheless,~a~single~plane wave is not
localised in space.~As a result,~$|\psi_{Y,P}\rangle$
is non-normalisable,~implying that $|\psi_{Y,P}\rangle$~is physically obscure in this
case.~This problem can be re-solved by treating~their~(normalisable) superposition~\cite{Itzykson&Zuber}:
\beqa\label{eq:non-covariant-wave-function}
\big(\psi_{Y,P}(y)\big)_\textrm{non-spinorial} &=& {\int}\frac{d^4K}{(2\pi)^3}\,
\theta\big(K^0\big)\,\delta\big(K^2-M^2\big)\,F_P(K)\,u(K)\,e^{-iK{\cdot}(y - Y)}\,,
\eeqa
where $F_P(K)$ has a narrow peak at $K = P$.~Yet,~unlike the
plane~wave~\eqref{eq:non-normalisable-wave-function},~this wave function does not properly transform
under the Lorentz transformations.~In fact,~it~is not~a~spinor.~For this reason, we instead consider in
what follows that
\beqa\label{eq:wave-function}
\psi_{Y,P}(y) &\equiv& {\int}\frac{d^4K}{(2\pi)^3}\,
\theta\big(K^0\big)\,\delta\big(K^2-M^2\big)\,
F_P(K)\,\frac{\gamma{\cdot}K + M}{2M}\,u(P)\,e^{-iK{\cdot}(y - Y)}\,,
\eeqa
where $\gamma^a$ are the four Dirac matrices in Weyl's representation~\cite{Peskin&Schroeder}.~This
wave function can~be shown to be~a~solution of the Dirac equation and to transform as a spinor
under the Lorentz transformations, assuming $F_P(K)$ is a Lorentz scalar (see below).

As mentioned above,~one normally treats on-mass-shell plane waves in theoretical~particle
physics, which are associated with in- and out-going particles in a given scattering amplitude.
This amplitude depends on initial and final momenta of such particles, but not~on
their initial and final positions.~Such amplitude does not disappear~if~the~particles~are
infinitely separated away from each other.~This is in contradiction to observations.~In theory,\,this
circumstance~is taken into account via the cluster decomposition principle
(see Ch.~4 in~\cite{Weinberg}),~basically stating that distant scattering experiments yield
uncorrelated results.~For this principle to hold,~one needs
localised-in-space quantum states which correspond to wave packets.

Wave packets involve, at least, one additional parameter to determine the shape~of~$F_P(K)$.
This parameter appears in the Heisenberg uncertainty relation and is known in the quantum-
mechanics literature as momentum variance~\cite{Merzbacher}.~It ensures that
\beqa\nonumber
\langle \psi_{Y,P}|\psi_{Y,P}\rangle
&=& {\int}d^3\mathbf{y}\,(\psi_{Y,P}(y))^\dagger\,\psi_{Y,P}(y)
\\[1mm]\label{eq:normalisation-condition-minkowski}
&=& \frac{1}{2}{\int}\frac{d^3\mathbf{K}}{(2\pi)^3}\,
\frac{|F_{P}(\mathbf{K})|^2}{\sqrt{\mathbf{K}^2 + M^2}}\,\frac{M^2+P{\cdot}K}{2M^2} \;\equiv\; 1\,.
\eeqa
The right-hand side is Lorentz-invariant.~It requires 
the momentum integral be also invariant under the Lorentz transformations.
In other words, we must consider $F_P(K) = F(P{\cdot}K)$.~We have dealt with
Lorentz-invariant Gaussian wave functions in~\cite{Emelyanov-2020,Emelyanov-2021,Emelyanov-2022},
which have been earlier studied in~\cite{Naumov&Naumov,Naumov}.~We
wish here to treat a modified Lorentz-invariant Gaussian wave packet:
\beqa\label{eq:wave-function-profile}
F_{P}(K) &\equiv& \frac{2^{3/2}\pi M}{D\,\sqrt{K_1\hspace{-1.0mm}\left(\frac{M^2}{D^2}\right)}}\,
\frac{\exp\Big({-}\scalebox{1.3}{$\frac{P{\cdot}K}{2D^2}$}\Big)}{\sqrt{M^2+P{\cdot}K}}\,,
\eeqa
where $D$ is the momentum variance and $K_\nu(z)$ stands for the modified Bessel function
of~the second kind.~The pre-factor in~\eqref{eq:wave-function-profile} has been chosen for
the normalisation condition~\eqref{eq:normalisation-condition-minkowski}~to~be fulfilled
(cf.~\cite{Emelyanov-2020,Emelyanov-2021}). 

\subsection{Minkowski-spacetime approximation}

We have considered so far a spin-half wave packet in Minkowski spacetime.~The~Universe~is a
non-Minkowski spacetime~\cite{Mukhanov}.~Yet, it follows from Einstein's equivalence
principle~that~the observable Universe can be locally approximated by Minkowski spacetime
at any point~$Y$~for $y$ satisfying
\beqa
|y-Y| &\ll& l_c(Y)\,,
\eeqa
where $l_c(Y)$ is a local curvature length at $Y$,~estimated by the inverse of the fourth root~of~the
Kretschmann scalar at that point.~At the Earth's
surface,~we obtain~$l_c(r_\oplus) \sim 10^{11}\,\text{m}$,~meaning that the Earth's curvature can
be basically ignored in collider physics.~We~want~to~go~beyond
this approximation.~In other words,~\eqref{eq:wave-function} turns into a leading-order 
approximation of the~wave function which does not involve metric derivatives:
\bsubeqs
\beqa
\psi_{Y,P}^{(0)}(y) &=& {\int}\frac{d^4K}{(2\pi)^3}\,
\theta\big(K^0\big)\,\delta\big(K^2-M^2\big)\,F_P(K)\,\psi_{Y,P|K}^{(0)}(y)\,,
\eeqa
where we have on the mass shell that
\beqa\label{eq:wave-function-0}
\psi_{Y,P|K}^{(0)}(y) &=& 
\frac{\gamma{\cdot}K + M}{2M}\,u(P)\,e^{-iK{\cdot}(y - Y)}\,.
\eeqa
\esubeqs
It should be pointed out that
the momentum integral can be exactly evaluated with~the~choice \eqref{eq:wave-function-profile}
for $F_P(K)$, which is given by a combination of elementary functions.

The wave function is given in terms of local Minkowski (Riemann) coordinates $y$ defined~at
$Y$.~It can be re-written in general coordinates $x = x(y)$ with $X=x(Y)$.~In these~coordinates,
$\psi_{X,P}^{(0)}(x)$ depends on $x$ via $\sigma(x,X)$, $P^M(X)\,\sigma_M(x,X)$ and
$\gamma^M(X)\,\sigma_M(x,X)$,~where $\sigma(x,X)$~is a geodetic distance --
Synge's world function, -- $\sigma_M(x,X)$ is its derivative with respect~to~$X^M$,
$P^M(X) \equiv e_a^M(X)\,P^a$ and $\gamma^M(X) \equiv e_a^M(X)\,\gamma^a$,~where
$e_a^M(X)$ are vierbein fields at $X$,~namely $g_{MN}(X)\,e_a^M(X)\,e_b^N(X) = \eta_{ab}$.\,These
all mean that the wave packet transforms as a zero-rank tensor under
general coordinate transformations,~as required.

\subsection{Beyond Minkowski-spacetime approximation}
\label{sec:bmsta}

We now wish to obtain the leading-order curvature correction to~$\psi_{X,P}^{(0)}(x)$.~The
Dirac-field equation~generically~reads
\beqa\label{eq:dirac-field-equation}
\big(i\gamma^\mu(x) D_\mu - M\big)\,\psi_{X,P}(x) &=& 0\,,
\eeqa
where $D_\mu$ is the spinorial covariant derivative.~By
setting $Y = 0$ in what follows for the~sake~of simplicity,
we obtain in normal Riemann coordinates that
\bsubeqs
\beqa\label{eq:zeroth-order-ffeq}
\big(i\gamma^a \partial_a - M\big)\psi_{Y,P}^{(0)}(y) &=& 0\,,
\\[2mm]\label{eq:first-order-ffeq}
\big(i\gamma^a \partial_a - M\big)\psi_{Y,P}^{(2)}(y) &=& -\frac{i}{6}\,
R_{\;\;cbd}^{a}\,y^c y^d\gamma^b\partial_a\psi_{Y,P}^{(0)}(y) -
\frac{i}{8}\,R_{abcd}\,\gamma^c\gamma^b\gamma^a y^d\psi_{Y,P}^{(0)}(y)\,.
\eeqa
\esubeqs
Note,~$\psi_{Y,P}^{(2)}(y)$ depends apparently on the
curvature tensor~at~$Y$, where the index ``$(2)$"~refers to the number of metric
derivatives involved,~while $\psi_{Y,P}^{(1)}(y)$
does not exist in normal~Riemann coordinates.

In vacuum,~i.e.~$R_{ab} = 0$,~the second term on the right-hand side of~\eqref{eq:first-order-ffeq}
is identically~zero. In this case, we have
\bsubeqs
\beqa
\psi_{Y,P}^{(2)}(y) &=& {\int}\frac{d^4K}{(2\pi)^3}\,
\theta\big(K^0\big)\,\delta\big(K^2-M^2\big)\,F_P(K)\,\psi_{Y,P|K}^{(2)}(y)\,,
\eeqa
where we find on the mass shell that
\beqa\label{eq:wave-function-2}
\psi_{Y,P|K}^{(2)}(y) &=& \mathcal{O}^{(2)}\psi_{Y,P|K}^{(0)}(y)\,,
\eeqa
where by definition
\beqa\label{eq:o-2}
\mathcal{O}^{(2)} &\equiv&
\frac{iK{\cdot}y}{6M^2}\,R_{acbd}\,K^aK^by^cy^d
+\frac{1}{12M}\,R_{acbd}\,K^ay^cy^d\gamma^b
\nonumber\\[2mm]
&& +\,\frac{K{\cdot}y+ i}{12M^3}\,R_{acbd}\,K^aK^by^d\gamma^c
-\frac{2iK{\cdot}y+1}{8M^2}\,R_{acbd}\,K^ay^cS^{bd}\,,
\eeqa
and
\beqa\label{eq:s-tensor}
S^{ab} &\equiv& \frac{i}{4}\,\big[\gamma^a\,,\gamma^b\big]\,.
\eeqa
\esubeqs
This solution is non-unique,~like in case of the scalar-field model~\cite{Emelyanov-2021}.~For
instance,~we~obtain others by adding multiples of $R_{acbd}\,K^ay^c(K^by^d - S^{bd})$ and
$R_{acbd}\,K^a(K^by^d\gamma^c - iy^cS^{bd})$ to $\mathcal{O}^{(2)}$.
In this article,~however,~we shall focus our study on $\mathcal{O}^{(2)}$~as given in~\eqref{eq:o-2}.

Among of all terms entering $\mathcal{O}^{(2)}$,~only the first term modifies the
wave-function~phase~in~a gravitational field.~This term coincides with that we have found for
spin-zero particles~at~the leading order in space-time curvature~\cite{Emelyanov-2021}.~For this
reason,~spin should not influence~quantum interference induced by space-time curvature.~In
fact,~the relative phase shift of~wave~packets obtained via splitting an ultra-cold ${}^{87}\text{Rb}$
atom cloud~\cite{Asenbaum&etal} is oblivious to atoms' spin degree~of freedom.

\section{Observables}

\subsection{Quantum particle}

The quantum-particle state $|\psi_{Y,P}\rangle$ is defined in Minkowski spacetime through the creation
operator~\eqref{eq:particle-creation-minkowski}.~We covariantly generalise it to curved spacetime as follows:
\beqa
|\psi_{X,P}\rangle &\equiv& \hat{a}^\dagger(\psi_{X,P})|\Omega\rangle\,,
\eeqa
where
\beqa\label{eq:particle-creation-generic}
\hat{a}^\dagger(\psi_{X,P}) &\equiv&
{\int_\Sigma}d\Sigma_\mu(x)\,\hat{\overline{\Psi}}(x)\,\gamma^\mu(x)\,\psi_{X,P}(x)\,,
\eeqa
where $\Sigma$ is a time-like Cauchy surface on which,~however,~the integral does not
depend.~This~is because both the Dirac field and the wave packet are solutions of the
Dirac-field equation~\eqref{eq:dirac-field-equation} and $\psi_{X,P}(x)$ is localised in
space.~Next,~making use of the anti-commutation relation~for~$\hat{\Psi}(x)$ and its canonical
conjugate $\hat{\Pi}(x)$ and taking into account $\hat{a}(\psi_{X,P})|\Omega\rangle = 0$, we obtain
\beqa
\langle \psi_{X,P}|\psi_{X,P}\rangle &=& {\int_\Sigma}d\Sigma_\mu(x)\,J^\mu(x)\,,
\eeqa
where
\beqa\label{eq:current}
J^\mu(x) &\equiv& \overline{\psi}_{X,P}(x)\,\gamma^\mu(x)\,\psi_{X,P}(x)
\eeqa
is covariantly conserved,~namely~$\nabla_\mu J^\mu(x) = 0$.~It is worth
emphasising~that~$J^\mu(x)$ is a vector, because $\psi_{X,P}(x)$ is a scalar with respect to
general coordinate transformations.~This~covariant conservation law means that
$\langle \psi_{X,P}|\psi_{X,P}\rangle$~is~a constant which we set to unity:
\beqa\label{eq:normalisation-condition}
\langle \psi_{X,P}|\psi_{X,P}\rangle &\equiv& 1\,.
\eeqa
This defines the normalisation condition for the wave function $\psi_{X,P}(x)$ in curved spacetime.
Note that both sides of~\eqref{eq:normalisation-condition} are diffeomorphism
invariant.~This physically leads to~the~frame-independent existence of quantum particles
modelled by $|\psi_{X,P}\rangle$,~even in the presence of~a~non-stationary gravitational
field (see~\cite{Emelyanov-2020} for the de-Sitter-universe case).

This circumstance is at odds with the idea that quantum particles may be created~in~non-stationary
spacetimes~\cite{Birrell&Davies,Parker&Toms}.~This relies on a few assumptions.~For the
sake of concreteness, we focus on the flat de-Sitter spacetime,~i.e.~on that patch of the de-Sitter
hyperboloid,~which can be parametrised by Friedmann--Robertson--Walker coordinates with no spatial
curvature. In this example,~one first expands $\hat{\Psi}(x)$ over modes which are eigenfunctions
of Killing~vectors generating translations in space.~In the
absence of cosmic-time-translation symmetry,~there~is no unique choice of modes' dependence
on cosmic time.~For this reason,~one second imposes~a condition on how the modes have to depend
on time at past cosmic infinity, defining~$\psi_{-\infty}(x)$. One instead chooses
$\psi_{+\infty}(x)$ for that condition to be fulfilled at future cosmic~infinity as~well. Both
$\psi_{-\infty}(x)$ and $\psi_{+\infty}(x)$ can be utilised to define wave functions to satisfy the
normalisation condition~\eqref{eq:normalisation-condition} with $\psi_{X,P}(x)$ replaced
by $\psi_{-\infty}(x)$ and $\psi_{+\infty}(x)$, respectively.~The cosmological particle creation
is owing to the \emph{replacement} of $\psi_{-\infty}(x)$ by~$\psi_{+\infty}(x)$~\cite{Schroedinger}.~In
contrast,~$\psi_{X,P}(x)$ locally reduces to the plane-wave superposition in the vicinity of
\emph{any} $X$, which is required~for quantum particle theory to be consistent with
Einstein's equivalence principle~\cite{Emelyanov-2020,Emelyanov-2021,Emelyanov-2022}.

Another novel aspect is that~$|\psi_{X,P}\rangle$~is diffeomorphism invariant.~For example,~a part
of the flat de-Sitter spacetime can be parametrised by static coordinates.~In these coordinates,~there
is a time-like Killing vector which generates translations in time.~This basically eliminates~the
ambiguity in the mode choice by selecting those which are its eigenfunctions.~This property~is
correspondingly fulfilled for any moment of static-time coordinate.~In fact, such modes define
a quantum-particle state being unitarily equivalent to $|\psi_{-\infty}\rangle$.~At future cosmic
infinity,~$|\psi_{+\infty}\rangle$ is replaced by $|\psi_{-\infty}\rangle{\otimes}\dots{\otimes}|\psi_{-\infty}\rangle$
under the coordinate transformation from the flat to~static patch of the de-Sitter hyperboloid.~This
\emph{replacement} is apparently in tension with~the~general principle of relativity.

In the absence of experimental data favouring the idea of quantum particles being created in a
non-stationary spacetime geometry, it is unclear whether this model accurately describes 
particle physics in curved spacetime.~From other side,~the Einstein equivalence principle~is~by
now well tested in various experiments~\cite{Will},~while general covariance is a guiding principle~for
formulating physical laws.~These laws manifest themselves through interaction of (quantum)
particles.~Thus,~general covariance makes physical sense if and only if
(quantum) particles~are independent on coordinate reference frames.

In interacting quantum field theories,~$|\psi_{X,P}\rangle$ should be interpreted as an asymptotic
state to model either~an in- or out-coming particle.~In the former instance,~$|\psi_{X,P}\rangle$ does
change~with time.~It is because $\psi_{X,P}(x)$ is a solution of the
linear Dirac-field~equation, while $\hat{\Psi}(x)$ satisfies a non-linear Dirac-field
equation.~In fact,~we obtain from~\eqref{eq:particle-creation-generic} that
\beqa\label{eq:out-in}
\hat{a}^\dagger(\psi_{X,P})\big|_\text{out} &=& \hat{a}^\dagger(\psi_{X,P})\big|_\text{in} -i
{\int}d^4x\sqrt{-g(x)}
\big(
iD_\mu\hat{\overline{\Psi}}(x)\gamma^\mu(x) + M\hat{\overline{\Psi}}(x)
\big)\psi_{X,P}(x)\,,
\eeqa
where we have taken into account that $\psi_{X,P}(x)$ obeys~\eqref{eq:dirac-field-equation} and turns
to zero at spatial~infinity, i.e.\,$\psi_{X,P}(x)$ is localised in space.~This
diffeomorphism-invariant integral vanishes if and~only~if $\hat{\Psi}(x)$ satisfies
the linear Dirac-field equation.~The result~\eqref{eq:out-in} can be utilised to generalise~the
Lehmann--Symanzik--Zimmermann reduction~formula~\cite{Lehmann&Symanzik&Zimmermann,Srednicki}
to curved spacetime.~It is worth emphasising that
this generalisation logically follows from the general principle~of~relativity.
Hence,~the~quantum state
$|\psi_{X,P}\rangle$ may depend on time in a gravitational field
if we go~beyond classical gravity, by~working,~for instance,~in the framework of~the effective
quantum-gravity theory~\cite{Donoghue}.

Here we should digress to briefly discuss the quantum-vacuum decay in a constant electric-field
background to compare it with the vacuum decay in a non-stationary gravitational~field.\\
The former is known in the literature as Schwinger's effect which was
deduced in a manifestly gauge-invariant way by employing the proper-time
method~\cite{Schwinger}.~This has no explicit reference to quantum particles.~A (gauge-dependent)
derivation has been later proposed,~making use~of
the quantum-particle notion based on the Feynman prescription
distinguishing positive- and negative-frequency modes as,~respectively,~particles and
antiparticles~\cite{Nikishov},~similar~to~\cite{Schroedinger}.~In 
the electric-field presence,~$\hat{\Psi}(x)$ cannot generically satisfy the linear Dirac
equation,~meaning that the integral in~\eqref{eq:out-in} is
non-trivial,~unless $\psi_{X,P}(x)$ propagates away from the electric~field.
Furthermore,~to our knowledge,~a gauge-invariant wave-function solution 
to model~a~charged test particle
placed in a constant electric field is not yet known in the literature.

Working in normal Riemann coordinates, we have
\beqa
\langle \psi_{Y,P}|\psi_{Y,P}\rangle &\equiv& \langle \psi_{Y,P}|\psi_{Y,P}\rangle_{(0)} +
\langle \psi_{Y,P}|\psi_{Y,P}\rangle_{(2)} + \cdots\,,
\eeqa
where the first term is generically independent on metric derivatives,~whereas the
second~one depends on the curvature tensor at the point $Y$, i.e. on
no more than two metric derivatives. We find
in case of~\eqref{eq:wave-function-0} and~\eqref{eq:wave-function-2} that
\bsubeqs
\beqa
\langle \psi_{Y,P}|\psi_{Y,P}\rangle_{(0)} &=& 1\,,
\\[2mm]
\langle \psi_{Y,P}|\psi_{Y,P}\rangle_{(2)} &=& 0\,.
\eeqa
\esubeqs
These mean that the wave function $\psi_{X,P}(x) \approx \psi_{X,P}^{(0)}(x) + \psi_{X,P}^{(2)}(x)$
is properly normalised~up~to the leading-order approximation in the curvature tensor.

\subsection{Quantum-particle propagation}

We define particle's position through
the first moment of the current density~\eqref{eq:current}~\cite{Emelyanov-2020,Emelyanov-2021}:
\beqa\label{eq:position-expectation-value}
\langle x^\mu(\Sigma) \rangle &\equiv&
{\int_\Sigma}d\Sigma_\nu(x)\,x^\mu J^\nu(x)\,,
\eeqa
where $\Sigma$ is a time-like Cauchy surface.~Apparently, the position expectation value depends~on
the choice of $\Sigma$. In quantum mechanics, this choice determines the notion of time, which is
invariant under the Galilei transformations. In special relativity, this choice is ambiguous,~as
the Lorentz transformations mix temporal and spatial coordinates.~This~means~that~we~need
here a physical hypothesis which specifies preferred Cauchy surfaces.~We intend to~presume~in
this regard that quantum massive particles measure proper time which is~commonly~denoted
by $\tau$.~Despite of the fact that this assumption logically follows from the geodesic equation,~it
can also be justified by utilising experimental data.~Specifically,~muons~are~unstable~quantum
particles whose mean lifetime is about $2.19{\times}10^{-6}\,\text{s}$.~It implies that a
cosmic-ray muon~cannot reach the Earth's surface if its mean lifetime is measured by a clock
to rest~with~respect~to~the Earth.~Since this is at odds with observations, the clock and
muon time~differ~from~each~other. This discrepancy comes from the time-dilation
effect~\cite{Rossi&Hall}:~The laboratory lifetime~of~a cosmic-ray muon is by a Lorentz factor
bigger than its proper lifetime.

Apart from~\eqref{eq:position-expectation-value} reduces in the quantum-mechanics regime to the
well-known definition~of position expectation value, we obtain from it in the particle's rest frame
$\chi = (\tau,\boldsymbol{\chi})$ that
\beqa
\langle \dot{\chi}^a(\tau) \rangle &=& 
{\int}d^3 \boldsymbol{\chi}\,\sqrt{-g(\tau,\boldsymbol{\chi})}\,J^a(\tau,\boldsymbol{\chi})\,.
\eeqa
In Minkowski spacetime,~$\langle \dot{\chi}^i(\tau) \rangle$ identically vanishes.~Under the Lorentz
transformation~from the rest frame $\chi$ to $y$,~we then find
\beqa
\langle \dot{y}^a(\tau) \rangle &=& \Lambda_{\;\;b}^a\,\langle \dot{\chi}^a(\tau) \rangle
\;=\; U^a \quad \textrm{with} \quad U^a \;\equiv\; P^a/M\,,
\eeqa
as in classical theory.~Thus,~our definition of
quantum-particle position makes sense,~at~least in the absence of
gravity (cf. equation~(23) of Sec.~III.B in~\cite{Emelyanov-2020}).

Working in Riemann normal coordinates, we have
\beqa
\langle y^a(\tau) \rangle &\equiv& \langle y^a(\tau) \rangle_{(0)} + \langle y^a(\tau) \rangle_{(2)} + \cdots
\eeqa
where we find
\bsubeqs\label{eq:trajectory}
\beqa\label{eq:trajectory-0}
\langle y^a(\tau) \rangle_{(0)} &=& U^a\tau\,,
\\[2mm]\label{eq:trajectory-2}
\langle y^a(\tau) \rangle_{(2)} &=& -\frac{1}{8M}\,
\bigg(\frac{1}{4D^2}\,
{f_1}\hspace{-0.8mm}\left(\scalebox{1.2}{$\frac{M^2}{D^2}$}\right) +
{\tau^2}{{f_2}\hspace{-0.8mm}\left(\scalebox{1.2}{$\frac{M^2}{D^2}$}\right)}
\bigg)R_{\;\;bcd}^a\,U^b\langle S^{cd} \rangle\,,
\eeqa
\esubeqs
where by definition
\beqa\label{eq:spin-matrix-c}
\langle S^{ab} \rangle &\equiv& \bar{u}(U)S^{ab}u(U)\,,
\eeqa
where $S^{ab}$ has been defined in~\eqref{eq:s-tensor}, and
\bsubeqs
\beqa
f_1(z) &\equiv& 
7+\frac{11}{z} - \frac{55}{4z^2} + \text{O}\Big(\frac{1}{z^3}\Big),
\\[2mm]
f_2(z) &\equiv& 1-\frac{5}{2z} + \frac{75}{8z^2} + \text{O}\Big(\frac{1}{z^3}\Big).
\eeqa
\esubeqs
The result $\langle \ddot{y}^a(\tau) \rangle \neq 0$ implies that Dirac particles propagate
along non-geodesic trajectories, because, in classical theory, geodesics passing through
$Y = 0$ are given by straight world~lines in the Riemann frame~\cite{Petrov}.~Still,~this
effect comes from the spin degree of freedom.~Classical spinning particles are
known to be subject to Mathisson's force,~leading to their non-geodesic
motion in curved spacetime~\cite{Mathisson,Papapatrou} (see~\cite{Mashhoon} for a short
review).~To study this effect~further~in the framework of quantum field theory,
we~next intend to compute 4-momentum of spin-half quantum particles.

\subsection{Quantum-particle 4-momentum}
\label{sec:qpfm}

Making use of the anti-commutation relation for $\hat{\Psi}(x)$ and its canonical conjugate $\hat{\Pi}(x)$,
we obtain
\beqa
\langle p^\mu(\Sigma) \rangle &\equiv& {\int_\Sigma} d\Sigma^\nu(x)\,
\Big(\langle \psi_{X,P}|\hat{T}_\nu^\mu(x)|\psi_{X,P}\rangle - \langle \Omega|\hat{T}_\nu^\mu(x)|
\Omega\rangle\Big)
\nonumber\\[2mm]\label{eq:four-momentum}
&=& {\int_\Sigma} d\Sigma_\nu(x)\,\frac{i}{2}\,
\overline{\psi}_{X,P}(x)\gamma^{(\mu}(x)D^{\nu)}\psi_{X,P}(x) + \text{c.c.}\,,
\eeqa 
where we have explicitly subtracted the stress-tensor vacuum expectation value
$\langle \Omega|\hat{T}_\nu^\mu(x)|\Omega\rangle$ from
$\langle \psi_{X,P}|\hat{T}_\nu^\mu(x)|\psi_{X,P}\rangle$,
because that does not depend on the wave packet $\psi_{X,P}(x)$~and,~thus,
cannot provide a physical contribution to the particle energy-momentum tensor.~It should~be,
however, mentioned that $\langle \Omega|\hat{T}_\nu^\mu(x)|\Omega\rangle$ is commonly supposed to
have a physical meaning~\cite{Birrell&Davies}. Apparently,~this vacuum
expectation value is ill-defined because of the mathematical nature of
quantum-field operators.~A~\emph{properly} regularised and then
renormalised~$\langle \Omega|\hat{T}_\nu^\mu(x)|\Omega\rangle$~turns 
out to depend on the spacetime-curvature length
$l_c(x)$~\cite{Birrell&Davies},~which is finite due to the~presence of (quantum) matter (unless one
treats a purely de-Sitter
spacetime with $l_c(x) = \text{const}$).~If~so, $\langle \Omega|\hat{T}_\nu^\mu(x)|\Omega\rangle$
has to depend on wave packets of particles constituting (quantum) matter.~This appears to point
a logical flaw in the statement that the quantum~vacuum~$|\Omega\rangle$~is~a~no-particle state and
$\langle \Omega|\hat{T}_\nu^\mu(x)|\Omega\rangle$ makes a non-zero contribution to the energy
budget~of~the~observable Universe.~For this reason,~we suppose that
$\langle \psi_{X,P}|\hat{T}_\nu^\mu(x)|\psi_{X,P}\rangle - \langle \Omega|\hat{T}_\nu^\mu(x)|\Omega\rangle$
might enter~the semi-classical Einstein equation,~while a curvature
length arising from~that does not source~a non-zero stress-tensor vacuum expectation value for
other quantum fields from the Standard Model. 

Working in normal Riemann coordinates, we have
\beqa
\langle p^a(\tau) \rangle &\equiv&
\langle p^a(\tau) \rangle_{(0)} + \langle p^a(\tau) \rangle_{(2)} + \cdots\,,
\eeqa
where we find\footnote{It turns out that $\langle p^a(\tau) \rangle_{(2)}$ is
independent on $D/M$.~First,~it follows from direct computations~by assuming
$D/M$ is small.~We obtain no contributions up to the sixth order in perturbation~theory.~Second,
numerical computations show that $\langle p^a(\tau) \rangle_{(2)}$ is insensitive to various
values of $D/M \ge 1$.}
\bsubeqs
\beqa
\langle p^a(\tau) \rangle_{(0)} &=& {{f_3}\hspace{-0.8mm}\left(\scalebox{1.2}{$\frac{M^2}{D^2}$}\right)}
{P^a}\,,
\\[2mm]\label{eq:four-momentum-2}
\langle p^a(\tau) \rangle_{(2)} &=& -\frac{\tau}{4}\,
R_{\;\;bcd}^a\,U^b \langle S^{cd} \rangle\,,
\eeqa
\esubeqs
where by definition (see also Sec.~III.B.3 in~\cite{Emelyanov-2020})
\beqa
f_3(z) &\equiv& K_2(z)/K_1(z)\,.
\eeqa

As noted above, point-like spinning bodies are subject to Mathisson's
spin-curvature~force \cite{Mathisson,Papapatrou}.~We find from the first
Mathisson--Papapetrou equation (see~(18) in~\cite{Mashhoon}) by solving~it
in normal Riemann coordinates that
\beqa
p^a(\tau) &\approx& MU^a -\frac{\tau}{4}\,R_{\;\;bcd}^a\,U^b \langle S^{cd} \rangle\,,
\eeqa
where higher-order curvature terms have been omitted.~Hence,~$\langle p^a(\tau) \rangle$
reduces to the classical result in the limit $D/M \rightarrow 0$.~To our knowledge,~this was
first derived in quantum theory~in the framework of relativistic quantum
mechanics~in~\cite{Gorbatsievich}.~Later,~it was obtained~by~making~use of the WKB approximation in
\cite{Audretsch} (see also~\cite{Cianfrani&Montani}).~Recent quantum-mechanics results~for~spin-
half particles in curved spacetime can be found
in~\cite{Obukhov&Silenko&Teryaev}.~Recent
results gained
in the framework of the effective quantum theory of gravity show that the gravitational deflection
depends on quantum-particle spin~\cite{Bjerrum-Bohr&etal-2015-1,Donoghue&El-Menoufi,Bjerrum-Bohr&etal-2015-2}.

In classical theory, 4-momentum is proportional to 4-velocity for point-like particles.~Still,
in quantum theory over curved spacetime, we have from~\eqref{eq:trajectory} that
\beqa
\langle p^a(\tau) \rangle
&\approx&
M_i\,\langle \dot{y}^a(\tau) \rangle - 
\frac{\tau D^2}{4M^2}\,
R_{\;\;bcd}^a\,U^b \langle S^{cd} \rangle\,,
\eeqa
where the inertial mass $M_i$ has been defined through the Lagrangian mass $M$ as
follows~\cite{Emelyanov-2022}:
\beqa
M_i &\equiv& M{f_3}\hspace{-0.8mm}\left(\scalebox{1.2}{$\frac{M^2}{D^2}$}\right).
\eeqa
We recover the classical result for point-like particles in
the limit $D/M \rightarrow 0$,~which~means~that the wave function has a definite value of the
momentum,~but no definite position~in~space.~The latter can also be seen
in~\eqref{eq:trajectory-2},~which~makes~no~longer~physical sense in the limit of vanishing 
momentum variance.

To study this limit in more detail,~we want to consider~\eqref{eq:wave-function-0}
and~\eqref{eq:wave-function-2} with $K \rightarrow P$.~This corresponds to setting
the momentum variance $D$ to zero.~In this instance,~the wave~function is non-normalisable.~It
particularly means that we must re-consider~our~method~of~computing observables.~We have learned
above that $J^a(y)$ may be interpreted as 4-velocity density~of~the quantum particle.~Therefore,~we
assume for the moment that
\beqa\label{eq:4-velocity-guess-1}
v^a(y) &\propto& \overline{\psi}_{Y,P}(y)\,\gamma^a(y)\,\psi_{Y,P}(y)\,,
\eeqa
where the normalisation factor needs to be determined.~We then find
\bsubeqs
\beqa
v_{(0)}^a(y)\big|_{y \,=\, U\tau} &\propto& U^a\,\overline{u}(P)u(P)\,,
\\[2mm]
v_{(2)}^a(y)\big|_{y \,=\, U\tau} &=& 0\,.
\eeqa
\esubeqs
Apparently, the last result cannot be consistent with ours for $\langle \dot{y}(\tau)\rangle_{(2)}$.~It
was~proposed~in~the references~\cite{Audretsch,Audretsch-2},~however,~to define 4-velocity
in the WKB approximation as follows:
\beqa\label{eq:4-velocity-guess-2}
v^a(y) &\propto& \frac{1}{2Mi}
\big(D^a\overline{\psi}(y)_{Y,P}\,\psi_{Y,P}(y) - \overline{\psi}_{Y,P}(y)\,D^a\psi_{Y,P}(y)\big)\,,
\eeqa
giving
\bsubeqs
\beqa
v_{(0)}^a(y)\big|_{y \,=\, U\tau} &\propto& U^a\,\overline{u}(P)u(P)\,,
\\[2mm]
v_{(2)}^a(y)\big|_{y \,=\, U\tau} &\propto& -\frac{\tau}{2}\,MR_{\;\;bcd}^a\,U^b \langle S^{cd} \rangle\,.
\eeqa
\esubeqs
If the normalisation factor is $1/2M$,~then we obtain 
$\langle \dot{y}(\tau)\rangle_{(0)}$ and $\langle \dot{y}(\tau)\rangle_{(2)}$ with $D$~set~to~zero. 
According to Gordon's decomposition,~\eqref{eq:4-velocity-guess-1} differs
from~\eqref{eq:4-velocity-guess-2} by a term which is proportional to
$\nabla_b(\overline{\psi}(y)S^{ab}(y)\psi(y))$.~This contributes neither to
$\langle \dot{y}(\tau)\rangle_{(0)}$ nor to $\langle \dot{y}(\tau)\rangle_{(2)}$ if $D/M \rightarrow 0$,~but
is~non-vanishing if computed on $y = U\tau$. Furthermore, we find
\bsubeqs
\beqa
\big(v^b(y)\nabla_b v^a(y)\big)_{(0)}\big|_{y \,=\, U\tau} &=& 0\,,
\\[2mm]
\big(v^b(y)\nabla_b v^a(y)\big)_{(2)}\big|_{y \,=\, U\tau} &\propto&
-\frac{1}{2}\,MR_{\;\;bcd}^a\,U^b \langle S^{cd} \rangle\,,
\eeqa
\esubeqs
which are consistent with $\langle \ddot{y}(\tau)\rangle_{(0)}$ and
$\langle \ddot{y}(\tau)\rangle_{(2)}$, respectively, if $D/M \rightarrow 0$ is considered.

\subsection{Quantum-particle spin}

We define the spin matrix as follows:
\beqa
\langle s^{\mu\nu}(\Sigma) \rangle &\equiv& {\int_\Sigma} d\Sigma_\lambda(x)\,
\Big(\langle \psi_{X,P} |\hat{S}^{\lambda\mu\nu}(x)|\psi_{X,P}\rangle - \langle \Omega|\hat{S}^{\lambda\mu\nu}(x)|\Omega\rangle\Big)
\nonumber\\[2mm]\label{eq:spin-matrix}
&=& {\int_\Sigma} d\Sigma_\lambda(x)\,
\frac{1}{2}\,
\overline{\psi}_{X,P}(x)\{\gamma^\lambda(x),S^{\mu\nu}(x)\}\psi_{X,P}(x)\,,
\eeqa
where $\hat{S}^{\lambda\mu\nu}(x)$ denotes the spin-matrix
operator~\cite{Itzykson&Zuber,Bogolyubov&Shirkov}.

Working in normal Riemann coordinates, we have
\beqa
\langle s^{ab}(\tau) \rangle &\equiv&
\langle s^{ab}(\tau) \rangle_{(0)} + \langle s^{ab}(\tau) \rangle_{(2)} + \cdots\,,
\eeqa
where we find
\bsubeqs\label{eq:spin-precession-effect}
\beqa
\langle s^{ab}(\tau) \rangle_{(0)} &=& \frac{1}{2}\,
{{f_4}\hspace{-0.8mm}\left(\scalebox{1.2}{$\frac{M^2}{D^2}$}\right)} \langle S^{ab} \rangle\,,
\\[2mm]
\langle s^{ab}(\tau) \rangle_{(2)} &=& \frac{1}{12}\,
\bigg(\frac{1}{4D^2}\,
{f_5}\hspace{-0.8mm}\left(\scalebox{1.2}{$\frac{M^2}{D^2}$}\right) +
{\tau^2}{{f_6}\hspace{-0.8mm}\left(\scalebox{1.2}{$\frac{M^2}{D^2}$}\right)}\bigg)
\bigg(R_{\;\;ecd}^{[a}\,U^{b]}U^e -\frac{1}{2}\,R_{\;\;\;\;cd}^{ab}\bigg) \langle S^{cd} \rangle\,,
\eeqa
\esubeqs
where by definition
\bsubeqs
\beqa
f_4(z) &\equiv&
1 - \frac{1}{z} + \frac{9}{4z^2} + \text{O}\Big(\frac{1}{z^3}\Big)\,,
\\[2mm]
f_5(z) &\equiv& 1 -\frac{14}{z} + \frac{87}{2z^2}+\text{O}\Big(\frac{1}{z^3}\Big)\,,
\\[2mm]
f_6(z) &\equiv& 1 -\frac{1}{z} - \frac{1}{4z^2}+\text{O}\Big(\frac{1}{z^3}\Big)\,.
\eeqa
\esubeqs

First, note that both $\langle s^{ab}(\tau) \rangle_{(0)}$ and
$\langle s^{ab}(\tau) \rangle_{(2)}$ separately satisfy the Pirani condition~\cite{Pirani}
in the following form:
\bsubeqs
\beqa
U_b\langle s^{ab}(\tau) \rangle_{(0)} &=& 0\,,
\\[2mm]
U_b\langle s^{ab}(\tau) \rangle_{(2)} &=& 0\,.
\eeqa
\esubeqs
Nevertheless,~it seems that the condition might need to be re-defined in quantum theory,~such
that it is formulated as a single expectation value.~This issue goes beyond our purpose~in~this
article and, therefore, we leave it aside.

Second,~we~have from~\eqref{eq:spin-precession-effect} that
\beqa
\langle \dot{s}^{ab}(\tau) \rangle &\approx&
\frac{\tau}{6}\,\bigg(1-\frac{D^2}{M^2}\bigg)
\bigg(R_{\;\;ecd}^{[a}\,U^{b]}U^e-\frac{1}{2}\,R_{\;\;\;\;cd}^{ab}\bigg) \langle S^{cd} \rangle\,.
\eeqa
If we assume for the moment that there is no spin
precession in the sense of~\cite{Rumpf},~namely~in~a local inertial
frame with the origin at the particle's centre of mass, which falls with it freely~--
a Fermi frame~\cite{Manasse&Misner},~-- then~we obtain from the~second
Mathisson--Papapetrou equation (see~(19) in~\cite{Mashhoon}) by solving it in normal
Riemann~coordinates that
\beqa 
\dot{s}^{ab}(\tau) &\approx& \frac{\tau}{6}\,
\bigg(R_{\;\;ecd}^{[a}\,U^{b]}U^e-\frac{1}{2}\,R_{\;\;\;\;cd}^{ab}\bigg) \langle S^{cd} \rangle\,.
\eeqa
Note, the spin precession occurs in the Riemann frame, as the right-hand side of this
equation does not vanish.~It is due to the fact that the Riemann frame
represents a local inertial frame at the given initial point ($\tau = 0$), while the Fermi frame along
the free-fall trajectory~($\tau \geq 0$).
The comparison of $\langle \dot{s}^{ab}(\tau) \rangle$ with $\dot{s}^{ab}(\tau)$ shows that
spin precesses in quantum theory in curved spacetime~in~the Fermi frame,~even in the absence
of torsion~\cite{Audretsch-2},~as the right-hand~side of\phantom{11}
\beqa
\frac{d}{d\tau} \big(\langle s^{ab}(\tau) \rangle-s^{ab}(\tau)\big) &\approx&
-\frac{\tau D^2}{6M^2}
\bigg(R_{\;\;ecd}^{[a}\,U^{b]}U^e-\frac{1}{2}\,R_{\;\;\;\;cd}^{ab}\bigg) \langle S^{cd} \rangle
\eeqa
cannot generically vanish,\,unless $D/M \rightarrow 0$,\,as in classical theory.\,This circumstance
might~be of use to determine the ratio
$D/M$ for a given quantum spin-half particle in satellite-borne experiments,
if the quantum-particle model proposed above adequately describes fermions~in gravity.

\section{Covariant spin-half antiparticle}

\subsection{Quantum antiparticle}

Generalising the antiparticle creation operator (see Sec.~41 in~\cite{Srednicki}) to curved
spacetime,~we define
\beqa
\hat{b}^\dagger(\varphi_{X,P}) &\equiv&
{\int_\Sigma}d\Sigma_\mu(x)\,\overline{\varphi}_{X,P}(x)\,\gamma^\mu(x)\,\hat{\Psi}(x)\,,
\eeqa
where the antiparticle wave function reads
\beqa
\varphi_{X,P}(x) &\equiv& \hat{C}\,\psi_{X,P}(x) \;=\; -i\gamma^2\big(\psi_{X,P}(x)\big)^*\,,
\eeqa
where $\hat{C}$ is the charge conjugation operator~\cite{Peskin&Schroeder} and the star means
complex conjugation.

\subsection{Particle-antiparticle symmetry in gravity}

Our computations of the quantum-antiparticle observables show that antiparticles cannot be
distinguished from particles in a gravitational field.~Specifically,~we obtain the same~results for
$\langle y^a(\tau)\rangle$,~$\langle p^a(\tau)\rangle$ and $\langle s^{ab}(\tau)\rangle$ with
$u(U)$ replaced in~\eqref{eq:spin-matrix-c} by $v(U) = \hat{C}u(U)$.~This~means~the\\
particle-antiparticle symmetry in gravity,~because the charge conjugation operator
preserves spin orientation.

\section{Non-inertial-frame observables}

\subsection{Quantum corrections to free-fall trajectory}

We have~so~far computed the quantum-particle trajectory in normal Riemann coordinates.
From an experimental viewpoint,~it is of interest to obtain its trajectory relative~to~a~detector
being at rest with respect to a given reference frame.~We assume that this frame is described
by general coordinates, $x = (t,\mathbf{x})$.~We then obtain from~\eqref{eq:local-inertial-coordinates}
(with higher-order corrections~in metric derivatives included~\cite{Petrov})
and~\eqref{eq:position-expectation-value} that
\bsubeqs\label{eq:non-inertial-frame-trajectory}
\beqa\label{eq:non-inertial-frame-trajectory-a}
\langle x^\lambda(\tau) \rangle_{(0)} &=& X^\lambda + U^\lambda\tau\,,
\\[3mm]\label{eq:non-inertial-frame-trajectory-b}
\langle x^\lambda(\tau) \rangle_{(1)} &=& - \frac{1}{2}\,
\Gamma_{\mu\nu}^\lambda
\Big(U^\mu U^\nu\tau^2
+\frac{1}{3}\,\langle\boldsymbol{\chi}^2(\tau) \rangle\big(U^\mu U^\nu - g^{\mu\nu}\big)\Big)\,,
\\[2mm]
\langle x^\lambda(\tau) \rangle_{(2)} &=& - \frac{1}{6}
\Big(\Gamma_{\mu\nu,\rho}^\lambda - 2\Gamma_{\mu\sigma}^\lambda\Gamma_{\nu\rho}^\sigma\Big)
\Big(U^\mu U^\nu U^\rho \tau^3 + 
\tau\,\langle\boldsymbol{\chi}^2(\tau) \rangle\big(U^\mu U^\nu U^\rho - U^{(\mu}g^{\nu\rho)}\big)
\Big)
\nonumber\\[1mm]
&&-\,\frac{1}{8M}\,
\bigg(\frac{1}{4D^2}\,
{f_1}\hspace{-0.8mm}\left(\scalebox{1.2}{$\frac{M^2}{D^2}$}\right) +
{\tau^2}{{f_2}\hspace{-0.8mm}\left(\scalebox{1.2}{$\frac{M^2}{D^2}$}\right)}\bigg)
R_{\;\;\mu\nu\rho}^\lambda\,U^\mu \langle S^{\nu\rho} \rangle\,,
\eeqa
\esubeqs
where $U^\mu = e_a^\mu\,U^a$
and $\langle\boldsymbol{\chi}^2(\tau) \rangle$ describes wave-packet spreading~\cite{Merzbacher},
namely
\beqa
\frac{1}{3}\,\langle\boldsymbol{\chi}^2(\tau) \rangle &\equiv&
\frac{1}{4D^2}\,{f_7}\hspace{-0.8mm}\left(\scalebox{1.2}{$\frac{M^2}{D^2}$}\right)
+ {{f_8}\hspace{-0.8mm}\left(\scalebox{1.2}{$\frac{M^2}{D^2}$}\right)}\,\frac{D^2\tau^2}{M^2}\,,
\eeqa
where by definition
\bsubeqs
\beqa
f_7(z) &\equiv&
1 - \frac{1}{2z} + \frac{7}{8z^2} + \text{O}\Big(\frac{1}{z^3}\Big),
\\[2mm]
f_8(z) &\equiv& 1 - \frac{7}{2z} + \frac{123}{8z^2}+\text{O}\Big(\frac{1}{z^3}\Big).
\eeqa
\esubeqs
The free-fall trajectory is thus modified also due to the wave-packet
spreading~\cite{Emelyanov-2021,Emelyanov-2022}.~It~is~a universal phenomenon
in quantum theory.~The momentum variance $D$ enters the Heisenberg uncertainty
relation, meaning that $D > 0$.~As a result,~the universality~of~free~fall or, in other words,
the weak equivalence principle is at odds with Heisenberg's uncertainty principle.

\subsubsection{Earth's gravitational field}

The Earth's gravitational field may be approximately modelled by the line element
\beqa
ds_\oplus^2 &\approx& \left(1 - \frac{r_{S,\,\oplus}}{|\mathbf{x}|}\right)dt^2
-4\,\frac{{\mathbf{J}_\oplus}\hspace{-0.5mm}{\times}\mathbf{x}}{|\mathbf{x}|^3}{\cdot}d\mathbf{x}\,dt - 
\left(1 + \frac{r_{S,\,\oplus}}{|\mathbf{x}|}\right)d\mathbf{x}^2\,,
\eeqa
where $r_{S,\,\oplus}$ is the Schwarzschild radius of Earth and
$\mathbf{J}_\oplus$ is its angular momentum.~We obtain
from~\eqref{eq:non-inertial-frame-trajectory-a} and~\eqref{eq:non-inertial-frame-trajectory-b}
in the non-relativistic~limit
at the Earth's surface that
\beqa\label{eq:free-fall-trajectory}
M_i\langle \ddot{\mathbf{x}}(\tau) \rangle &\approx& -M_gg_\oplus\mathbf{n} -
2M_g(\boldsymbol{\omega}_\oplus{\times}\mathbf{V})
-3M_g\big((\mathbf{n}{\cdot}\boldsymbol{\omega}_\oplus)\,
\mathbf{n}{\times}\mathbf{V}-
\boldsymbol{\omega}_\oplus{\times}\mathbf{V}\big)\,,
\eeqa
where the gravitational mass $M_g$ has been defined through the Lagrangian mass $M$ as
follows:
\beqa
M_g &\equiv& M_i
\bigg(1 +{{f_8}\hspace{-0.8mm}\left(\scalebox{1.2}{$\frac{M^2}{D^2}$}\right)}\frac{D^2}{M^2}\bigg).
\eeqa
Furthermore,~$\mathbf{n}$ stands for the three-dimensional unit vector radially pointing outwards,~and
\beqa
\boldsymbol{\omega}_\oplus &\equiv& \frac{2\mathbf{J}_\oplus}{r_\oplus^3}\,,
\eeqa
where $r_\oplus$ is the Earth's radius and
\beqa
V^i &\equiv& P^i/P^0\,.
\eeqa
It should be noted that the third term on the right-hand~side
of~\eqref{eq:free-fall-trajectory} arises from the gradient of the Earth's angular velocity and vanishes
if $\boldsymbol{\omega}_\oplus \propto \mathbf{n}$.

\subsubsection{Uniformly accelerated and rotating frame}

The result~\eqref{eq:free-fall-trajectory} should next be compared with the Dirac-particle trajectory in a
uniformly accelerated and rotating frame:
\beqa
ds^2 &=& \big((1+\mathbf{a}{\cdot}\mathbf{x})^2 -(\boldsymbol{\omega}{\times}\mathbf{x})^2\big)dt^2
-2(\boldsymbol{\omega}{\times}\mathbf{x}){\cdot}d\mathbf{x}\,dt - d\mathbf{x}^2\,,
\eeqa
where $\mathbf{a}$ and $\boldsymbol{\omega}$ are, respectively, constant acceleration and angular
velocity.~We find~from~\eqref{eq:non-inertial-frame-trajectory} in~the non-relativistic limit that
\beqa\label{eq:free-fall-trajectory-uar}
M_i\langle \ddot{\mathbf{x}}(\tau) \rangle &\approx& -M_i\mathbf{a} -
2M_g(\boldsymbol{\omega}{\times}\mathbf{V})\,,
\eeqa
where we have only taken into account terms linearly depending on the acceleration and~the angular velocity.

If we treat the acceleration-dependent part of the results~\eqref{eq:free-fall-trajectory}
and~\eqref{eq:free-fall-trajectory-uar},~then~the~difference
between these is owing to gravitational-length contraction which~is,
apparently,~non-existent in the uniformly accelerated frame.~In other words,~$M_g \neq M_i$~cannot
be gained by considering the gravitational time dilation only.~The latter might be interpreted
as causing free fall~in~its standard form~\cite{Czrnecka&Czarnecki}.

If we take into account the angular velocity,~then we find that the Coriolis force enters~the
equation of motion with the gravitational mass $M_g$ instead of the inertial mass~$M_i$.~This~also
holds in a uniformly accelerated and rotating frame.~This may be of use to determine the~ratio $M_g/M_i$
for a quantum particle in such a frame.

Note, in general, the difference between $M_g$ and $M_i$ disappears in the quantum-mechanics
limit,~$c \rightarrow \infty$,~as this results in $D/Mc \rightarrow 0$.~The free-fall non-universality in
quantum~theory is in this sense a relativistic effect,~which is in agreement with~\cite{Laemmerzahl}.

\subsection{Quantum corrections to gravitational-potential energy}

We compute next the quantum-particle energy in the general coordinate frame $x = (t,\mathbf{x})$. Its
energy is given by $\langle p_t(\tau) \rangle$ which,~in general,~may depend on the proper time.~In
particular, assuming that $\partial_t$ is a Killing vector, we have from~\eqref{eq:local-inertial-coordinates}
(with higher-order corrections in metric derivatives included~\cite{Petrov}) and~\eqref{eq:four-momentum}
that
\bsubeqs
\beqa
\langle p_t(\tau) \rangle_{(0)} &=& e_t^a\langle p_a(\tau) \rangle_{(0)}\,,
\\[2mm]
\langle p_t(\tau) \rangle_{(1)} &=& \frac{1}{2}\,g_{t[\nu,\mu]}\,e_a^\mu e_b^\nu
\langle l^{ab}(\tau) \rangle_{(0)}\,,
\\[3mm]
\langle p_t(\tau) \rangle_{(2)} &=& 0\,,
\eeqa
\esubeqs
where $\langle l^{ab}(\tau) \rangle$ is the angular-momentum
matrix~\cite{Itzykson&Zuber,Bogolyubov&Shirkov},~defined as the skew-symmetric part~of
the first moment of the energy-momentum tensor.~We find
\beqa
\langle l^{ab}(\tau) \rangle_{(0)} = \frac{1}{2}\,\langle S^{ab} \rangle\,, 
\eeqa
which should be compared with the result for $\langle s^{ab}(\tau) \rangle_{(0)}$ derived above.~It
immediately~shows that $M_g$ cannot enter the quantum-particle energy in the leading order
of the approximation. The physical consequence of this result will be discussed shortly.

It should be emphasised that the curvature tensor does not contribute to the
gravitational-potential energy (at the leading order of perturbation theory).~This is a
consequence~of~a~non-trivial cancellation of $\langle p^a(\tau) \rangle_{(2)}$
in~\eqref{eq:four-momentum-2} by a term arising from the coordinate
transformation $x = x(y)$ at the corresponding order of perturbation theory.~This cancellation is
in agreement with the observation we have made in the end of Sec.~\ref{sec:bmsta} that
spin does~not~affect~phase~shift induced by the curvature tensor.

\subsubsection{Earth's gravitational field}

We find in the Earth's gravitational field at the Earth's surface that
\bsubeqs\label{eq:energy-earth}
\beqa
\langle p_t(\tau) \rangle_{(0)} &\approx& \gamma M_i\big(1 - g_\oplus r_\oplus\big)
-\gamma \boldsymbol{\omega}_\oplus{\cdot}\mathbf{L}\,,
\\[2mm]
\langle p_t(\tau) \rangle_{(1)} &\approx& - 
\boldsymbol{\omega}_\oplus{\cdot}\mathbf{S} +  \mathbf{S}{\cdot}(g_\oplus\mathbf{n}{\times}\mathbf{V})
-\frac{3}{2}\big((\mathbf{n}{\cdot}\boldsymbol{\omega}_\oplus)\,
\mathbf{n}{\cdot}\mathbf{S}-\boldsymbol{\omega}_\oplus{\cdot}\mathbf{S}\big)\,,
\eeqa
\esubeqs
where we have taken into account terms depending linearly on $\mathbf{g}$ and
$\boldsymbol{\omega}$
only,~which~explains the approximation sign used in both equations,~and,~by definition,~$\gamma$ is the
Lorentz factor~and
\bsubeqs
\beqa
L^i &\equiv& \epsilon^{ijk}X^j(M_iV^k)\,,
\\[2mm]
S^i &\equiv& \frac{1}{4}\,\epsilon^{ijk} \langle S^{jk} \rangle\,.
\eeqa
\esubeqs

Let us now suppose that a quantum neutron is initially placed at the altitude $Z$.~We~then
obtain from~\eqref{eq:energy-earth} in the non-relativistic limit that
\beqa
\langle E \rangle &\approx& \frac{1}{2}\,M_i\mathbf{V}^2 + M_ig_\oplus Z\,.
\eeqa
If we now assume that this
quantum particle is observed at a later moment of time ($\tau > 0$)~at
\beqa
\langle z \rangle &\approx& Z + V_Z\tau - \frac{1}{2}\,\frac{M_g}{M_i}\,g_\oplus \tau^2\,,
\eeqa
according~to~\eqref{eq:free-fall-trajectory},~then its energy after the measurement must be bigger
than its initial~value. It is owing to $M_g/M_i > 1$, namely the total-energy
gain equals $(M_g -M_i)g_\oplus Z$.~We~then~obtain for neutrons
from~\cite{McReynolds,Dabbs&Harvey&Paya&Horstmann} and~\eqref{eq:free-fall-trajectory} that
\beqa
\frac{D^2}{(Mc)^2}\bigg|_\textrm{neutron} & \lesssim& 10^{-3}\,.
\eeqa
Therefore,~a gravitational-spectrometer resolution must be better than $10^{-10}\,\text{eV}$
to probe~the universality of free fall at quantum level with a better accuracy than
in~\cite{McReynolds,Dabbs&Harvey&Paya&Horstmann}, if the free-fall altitude,~$Z$,~is about
$1\,\text{m}$.~A gravitational spectrometer of the type treated in~\cite{Kulin&etal} might
thus be of use to provide an independent experimental result testing the quantum-particle
model proposed here.

\subsubsection{Uniformly accelerated and rotating frame}

We find in the uniformly accelerated and rotating frame that
\bsubeqs\label{eq:energy-arf}
\beqa
\langle p_t(\tau) \rangle_{(0)} &\approx& \gamma M_i\big(1 + \mathbf{a}{\cdot}\mathbf{X}\big)
-\gamma \boldsymbol{\omega}{\cdot}\mathbf{L}\,,
\\[2mm]
\langle p_t(\tau) \rangle_{(1)} &\approx& - 
\boldsymbol{\omega}{\cdot}\mathbf{S} +  \mathbf{S}{\cdot}(\mathbf{a}{\times}\mathbf{V})\,.
\eeqa
\esubeqs
This result is to compare with that obtained in~\cite{Hehl&etal}, wherein the Hamilton operator
has been derived in the uniformly accelerated and rotating frame in the non-relativistic limit,
$|\mathbf{V}| \ll 1$. The spin-rotation effect, which is owing to
$-\boldsymbol{\omega}{\cdot}\mathbf{S}$, has been first proposed in~\cite{Mashhoon-1988}
for a neutron interferometer.~We also have the so-called inertial spin-orbit
coupling,~i.e.~$\mathbf{S}{\cdot}(\mathbf{a}{\times}\mathbf{V})$,~which~is seemingly~by
a factor of $2$ bigger than that found in~\cite{Hehl&etal}.~This can be
readily~accounted~for~the Foldy--Wouthuysen transformation used in~\cite{Hehl&etal} to derive
a non-relativistic approximation for the Hamilton operator in the framework of relativistic
quantum mechanics.~In quantum field theory,~we obtain the expectation values~\eqref{eq:energy-arf} 
in a single-particle state,~which are independent on any unitary transformation.

\section{Concluding discussion}

The Standard Model of elementary particle physics employs quantum
field theory~in~order to describe high-energy scattering processes generically involving a
non-conserved number~of particle
species~\cite{Weinberg,Peskin&Schroeder,Itzykson&Zuber,Bogolyubov&Shirkov,Srednicki}.~The
primary object in quantum field theory is a quantum field.~This~is~a local distribution-valued
operator defined over a certain spacetime.~The Poincar\'{e} isometry~of 
Minkowski spacetime plays a key role in relating quantum fields to elementary and composite
particles to acquire physical meaning well before and after scattering~processes.~Specifically,
the Lehmann--Symanzik--Zimmermann reduction formula
links free particles with asymptotic quantum states
defining $S$-matrix elements~\cite{Lehmann&Symanzik&Zimmermann,Srednicki}.

Gravity is still treated as physics beyond the Standard Model of elementary particles,~even
though the gravitational interaction is known by now for several centuries.~Indeed,~there~are
no data which would necessitate quantum gravity for their explanation -- gravitation~is~far~too
weak with respect to the rest fundamental interactions to be noticeable in colliders.~Classical
gravitational phenomena are,~however,~successfully described by general relativity,~providing
a geometrical description for the gravitational interaction,~based on a number of ideas~among
of which local Poincar\'{e} invariance of the laws of nature~and general covariance.~Matter~curves 
spacetime in general relativity,~meaning theoretical particle~physics~relies~on~the~Minkowski-
spacetime approximation.

Direct observations of free fall of neutrons at the Earth's surface~\cite{McReynolds,Dabbs&Harvey&Paya&Horstmann} show these~particles
fall down over classical geodesics,~or,~at least,~these experiments were not sensitive enough to
notice any deviation from those.~From a fundamental point of view,~these experiments require
that quantum field theory in Minkowski spacetime be replaced by that in curved~spacetime.~A
serious obstacle arises here from the circumstance that it is unclear how to model~elementary
and composite particles in the absence of the global Poincar\'{e} symmetry.

Algebra of quantum fields,~rather than their particular realisation in terms of creation~and
annihilation operators,~should be of underlying relevance~\cite{Haag}.~We have
proposed~in~\cite{Emelyanov-2020,Emelyanov-2021,Emelyanov-2022}~to pick up
operators from the quantum-field algebra over curved spacetime,~which~locally~reduce
to those which create free particles in Minkowski spacetime.~These operators are employed~in
quantum field theory in Minkowski space by deducing~the Lehmann--Symanzik--Zimmermann
reduction formula~\cite{Lehmann&Symanzik&Zimmermann,Srednicki}.~The selection is achieved~by
means of a bi-scalar that,~at least~in~the
weak-field limit,~provides a wave-function description for quantum~particles.~Mathematically,
this is accomplished by constructing a wave packet depending on the space-time point $x$~via
the geodetic distance $\sigma(x,X)$ -- Synge's world function~\cite{DeWitt-1965}, -- and
their covariant derivatives at $X$,\,which are contracted with curvature tensors at that point.\,This
naturally allows to get~a wave function which is locally given by a superposition of
positive-frequency plane waves~as in Minkowski spacetime,~and which is a zero-rank
tensor with respect to general coordinate transformations.~This basic~idea~has been applied
so far to model spin-zero quantum particles in
gravity~\cite{Emelyanov-2020,Emelyanov-2021,Emelyanov-2022}.

We have proposed herein a model to describe spin-half quantum particles in~curved~space-
time in the framework of quantum field theory.~Its novelty consists again in assuming~that~the
Einstein equivalence principle and general covariance hold for spin-half quantum particles.~It
is not a self-evident assumption,~because the mainstream approach in quantum field theory~in
curved spacetime is based instead on exploiting global isometry group of a given spacetime~for
expressing quantum-field operators through creation and annihilation
operators~\cite{DeWitt,Birrell&Davies,Parker&Toms}.~Since local Poincar\'{e} group and
global isometry group are, generically, not isomorphic to each other, it is unclear if
global isometry group should be favoured with respect to local Poincar\'{e}~group,
taking into account that it is the latter which plays a fundamental role in high-energy particle
physics~\cite{Weinberg,Peskin&Schroeder,Itzykson&Zuber,Bogolyubov&Shirkov,Srednicki}.~Furthermore,
general covariance is abandoned in this approach.~It~follows~from
favouring different mode functions for the definition of creation and annihilation operators~in
different patches of same~spacetime.~From~an~experimental~standpoint,~this~is~in~tension~with
the Bonse--Wroblewski experiment~\cite{Bonse&Wroblewski}.~Indeed,~the~observed~interference
pattern is induced there by acceleration which,~in the leading-order of approximation,~enters
wave-packet~phase in the form~$-iM\mathbf{a}{\cdot}\mathbf{x}_R\,t_R$,~where
$x_R = (t_R,\mathbf{x}_R)$ are Rindler coordinates.~Wave functions~being eigenfunctions of the
Killing vector $\partial_{t_R}$ of Minkowski spacetime cannot have this acceleration-
dependent correction to~$-iMc^2\,t_R$,~which~are,~however,~normally used to do quantum
particle physics in Rindler spacetime~\cite{Birrell&Davies}.~In accord with general
covariance,~plane waves re-written~in
Rindler coordinates do have that observed correction~\cite{Emelyanov-2021}.

The first ever experiment that has shown that quantum physics is affected by gravity~is~the 
Collela--Overhauser--Werner experiment~\cite{Colella&Overhauser&Werner}.~The observed
interference pattern~produced~via overlapping two beams of thermal neutrons,~propagating
at different altitudes with respect~to the Earth's surface, is due to the free-fall acceleration.~It
was shown in the Bonse--Wroblewski experiment, mentioned above, that an analogous
effect takes place in an accelerated reference
frame.~So,~uniform gravity and acceleration cannot be distinguished in quantum-interference
experiments~\cite{Nauenberg}.~The quantum-particle model proposed here is consistent with this
empirical result.~It is also consistent with the observed phase shift due to
space-time curvature~\cite{Asenbaum&etal},~see \cite{Emelyanov-2021},~as well as with
the gravitational Aharonov--Bohm effect recently~experimentally~probed
in~\cite{Overstreet&etal}.~This suggests that our model deserves further scrutiny.

Quantum optical communication makes use of photons as elementary information carriers
\cite{Gisin&Thew}.~It is required here to take into account gravitational-field
background~to~keep~track~of~its systematic influence\,on the information distortion by long-distance
quantum communication.
Due to rapid developments of satellite-based quantum 
communication~\cite{Rideout&etal,Vallone&etal,Yin&etal},~it is necessary
to acquire a comprehensive insight into how photons are affected by the gravitational~field~of
Earth.~A model describing massless quantum particles needs to be established, going beyond
the semi-classical approximation by relying on such ideas as Born's statistical interpretation of
quantum measurements, general covariance and local Poincar\'{e} invariance.

The Earth's gravitational field is weak -- the local curvature length at the Earth's~surface~is
of the order of the astronomical unit.~It implies that the Minkowski-spacetime approximation 
used in elementary particle physics is adequate for the description of microscopic processes, 
whereas gravity must be taken into account over macroscopic time intervals.~Still,~it has~been argued~in~\cite{Zeldovich&Novikov,Hawking} that black holes of a microscopic size
might have been created in the~high- density state~of the early Universe.~The local curvature
length nearby such black holes~can~be much smaller than the hydrogen-atom size.~Thus,~the
question arises as to whether the wave-\\ packet description of elementary and composite
particles is adequate once their~quantum~size is comparable to a local curvature
length.~If affirmative,~it~must be clarified~what
the physical impacts~of~a~strong gravitational field on quantum matter are.

We have partially studied these questions in de-Sitter spacetime~\cite{Emelyanov-2020}:~A
coordinate-frame-
independent non-perturbative (in curvature) wave-packet solution deduced there
reveals that the kinematic properties of this wave solution noticeably differ from geodesic motion,
unless the inverse Hubble constant (curvature length) is much bigger than the wave-function
extent (position variance) which in turn must be much bigger than the Compton or de-Broglie~wave-
length of spin-zero quantum particles.~Preliminary computations show that there might~exist
a non-perturbative wave-packet solution in Einstein static universes.~Notably,~the~space-time
geometry inside an extremely slowly collapsing dust star~\cite{Oppenheimer&Snyder}~can be
approximated~by~a~closed Einstein~universe.~The difficulty consists in the circumstance
that the wave function depends, generically, on four scalars there,~while,~in
de-Sitter spacetime, on two scalars~\cite{Emelyanov-2020}.

The Newton equivalence principle states that gravitational and inertial mass of a body~are
equal in the non-relativistic and weak-gravity limit~\cite{Casola&etal}.~This principle is a result~of
numerous empirical tests and appears to be entirely accidental from a theoretical standpoint.~Still,~any
experiment has a limited degree of accuracy.~Furthermore,~quantum field theory and
general relativity are the most fundamental theories utilised nowadays
for the description of matter and gravitation.~Quantum theory of both matter and gravity
should be capable of~addressing the question how underlying the Newtonian principle actually
is.~Even though we have found $M_g \neq M_i$, the definition of $M_g$ does not follow from
the computation of Newton's gravitation potential sourced by the quantum particle.~This
computation requires to go beyond the test-
particle approximation.~In this sense,~our model is incomplete,~because the quantum-particle state
$|\psi_{X,P}\rangle$ is oblivious to gravity-field
operators,~e.g.~$\langle\psi_{X,P}|\hat{h}_{\mu\nu}(x)|\psi_{X,P}\rangle = 0$,~where
$\hat{h}_{\mu\nu}(x)$ is the graviton-field operator defined in the framework of
the effective field theory~of quantum gravity~\cite{Donoghue}.~Appropriately dressing
$\hat{a}^\dagger(\psi_{X,P})$ by an operator depending on $\hat{h}_{\mu\nu}(x)$ in the sense
of~\cite{Dirac} should give a way to determine active gravitational mass of a quantum particle.~There
is \emph{a priori} no guarantee that it matches passive gravitational mass, $M_g$.

A closely related issue is the (main) cosmological constant
problem~\cite{Pauli,Zeldovich,Weinberg-1989} following from the
\emph{assumption}~that~the quantum vacuum $|\Omega\rangle$ gravitates.~Indeed,
the semi-classical Einstein field~equation in vacuum reads
\beqa\label{eq:efq}
R_{\mu\nu}(x) - \frac{1}{2}\,R(x)\,g_{\mu\nu}(x) &=& 
\frac{8\pi G}{c^4}\, \langle \Omega |\hat{T}_{\mu\nu}(x)|\Omega\rangle\,,
\eeqa
where,~strictly
speaking,~$\langle \Omega| \hat{T}_{\mu\nu}(x)|\Omega\rangle$ is
divergent~\cite{Zeldovich,Weinberg-1989},~giving~thus~rise~to a non-physical
space-time geometry.~However,~this problem is a result of the
semi-classical approximation~in the sense that the metric tensor $g_{\mu\nu}(x)$
is obtained by solving~\eqref{eq:efq},~rather than by computing 
$\langle \Omega |\hat{g}_{\mu\nu}(x)|\Omega\rangle$.~For the semi-classical Einstein
field~equation~\eqref{eq:efq} to hold,~the
quantum vacuum $|\Omega\rangle$ must,~at least,~depend on $\hat{g}_{\mu\nu}(x)$ and
$\hat{T}_{\mu\nu}(x)$.~This can be achieved by dressing the vacuum state by an operator
depending on both $\hat{g}_{\mu\nu}(x)$ and $\hat{T}_{\mu\nu}(x)$.~At the~moment,~it~is
unclear~if~this is~logically~and physically admissible.

\section*{%\hspace*{-4.5mm}
ACKNOWLEDGMENTS}
It is a pleasure to thank A.K.\,Gorbatsievich for sharing with me the first reference~in~\cite{Gorbatsievich}.
I would also like to thank G.V.\,Kulin for the clarification of some aspects of~\cite{Kulin&etal}.


\begin{thebibliography}{99}

\bibitem{McReynolds}
A.W.\,McReynolds,
Phys. Rev. {\bf 83} (1951) 172.

\bibitem{Dabbs&Harvey&Paya&Horstmann}
J.W.T.\,Dabbs, J.A.\,Harvey, D.\,Paya, H.\,Horstmann,
Phys. Rev. {\bf 139} (1965) B756.

\bibitem{Koester}
L.\,Koester,
Z. Phys. {\bf 198} (1967) 187; Phys. Rev. D {\bf 14} (1976) 907.

\bibitem{Weinberg}
S.\,Weinberg,
\emph{Quantum Theory of Fields}
(Cambridge UP, 1995).

\bibitem{Peskin&Schroeder}
M.E.\,Peskin and D.V.\,Schroeder,
\emph{An Introduction to Quantum Field Theory}
(Addison-Wesley Publishing Company,~1995).

\bibitem{Itzykson&Zuber}
C.\,Itzykson and J.-B.\,Zuber,
\emph{Quantum Field Theory}
(McGraw-Hill Inc., 1980).

\bibitem{Bogolyubov&Shirkov}
N.N.\,Bogolyubov, D.V.\,Shirkov,
\emph{Quantum Fields}
(Benjamin-Cummings, Inc., 1983).

\bibitem{Srednicki}
M.\,Srednicki,
\emph{Quantum Field Theory}
(Cambridge UP, 2007).

\bibitem{DeWitt}
B.S.\,DeWitt,
Phys. Rep. {\bf 19} (1975) 295.

\bibitem{Birrell&Davies}
N.D.\,Birrell, P.C.W.\,Davies,
\emph{Quantum Fields in Curved Space}
(Cambridge UP, 1982).

\bibitem{Parker&Toms}
L.\,Parker, D.\,Toms,
\emph{Quantum Field Theory in Curved Spacetime} (Cambridge UP, 2009).

\bibitem{Will}
C.M.\,Will, Living Rev. Relativ. {\bf 17} (2014) 4.

\bibitem{Petrov}
A.Z.\,Petrov,
\emph{Einstein Spaces}
(Pergamon Press Ltd., 1969).

\bibitem{Colella&Overhauser}
R.\,Colella, A.W.\,Overhauser, 
Phys. Rev. Lett. {\bf 33} (1974) 1237.

\bibitem{Colella&Overhauser&Werner}
R.\,Colella, A.W.\,Overhauser, S.A.\,Werner, 
Phys. Rev. Lett. {\bf 34} (1975) 1472.

\bibitem{Bonse&Wroblewski}
U.\,Bonse, T.\,Wroblewski,
Phys. Rev. Lett. {\bf 51} (1983) 1401.

\bibitem{Nauenberg}
M.\,Nauenberg,
Am. J. Phys. {\bf 84} (2016) 879.

\bibitem{Emelyanov&Klinkhamer}
V.A.\,Emelyanov, F.R.\,Klinkhamer,
Acta Phys. Pol. B {\bf 52} (2021) 805.

\bibitem{Emelyanov-2020}
V.A.\,Emelyanov, Eur. Phys. J. C {\bf 81} (2021) 189.

\bibitem{Emelyanov-2021}
V.A.\,Emelyanov, Eur. Phys. J. C {\bf 82} (2022) 318.

\bibitem{Emelyanov-2022}
V.A.\,Emelyanov, Annalen der Physik {\bf 535} (2023) 2200386.

\bibitem{Einstein}
A.\,Einstein,
Annalen der Physik {\bf 49} (1916) 769.

\bibitem{Merzbacher}
E.\,Merzbacher,
\emph{Quantum Mechanics} (3rd Edition, John Wiley \& Sons, Inc., 1998).

\bibitem{Naumov&Naumov}
D.V.\,Naumov, V.A.\,Naumov,
J. Phys. G: Nucl. Part. Phys. {\bf 37} (2010) 105014.
 
 \bibitem{Naumov}
D.V.\,Naumov,
Phys. Part. Nuclei Lett. {\bf 10} (2013) 642.

\bibitem{Mukhanov}
V.F.\,Mukhanov,
\emph{Physical Foundations of Cosmology}
(Cambridge UP, 2005).

\bibitem{Asenbaum&etal}
P.\,Asenbaum \emph{et al.},
Phys. Rev. Lett. {\bf 118} (2017) 183602.

\bibitem{Schroedinger}
E.\,Schr\"{o}dinger,
Physica {\bf 6} (1939) 899.

\bibitem{Lehmann&Symanzik&Zimmermann}
H.\,Lehmann, K.\,Symanzik, W.\,Zimmermann, Nuovo Cimento {\bf 1} (1955) 205.

\bibitem{Donoghue}
J.F.\,Donoghue, Phys. Rev. D {\bf 50} (1994) 3874.

\bibitem{Schwinger}
J.\,Schwinger, Phys. Rev. {\bf 82} (1951) 664.

\bibitem{Nikishov}
A.I.\,Nikishov, Zh. Eksp. Teor. Fiz. {\bf 57} (1970) 1210 (Sov. Phys. JETP {\bf 30} (1970) 660).

\bibitem{Rossi&Hall}
B.\,Rossi, D.B.\,Hall, 
Phys. Rev. {\bf 59} (1941) 223.

\bibitem{Mathisson}
M.\,Mathisson,
Acta Phys. Pol. {\bf 6} (1937) 163.

\bibitem{Papapatrou}
A.\,Papapetrou,
Proc. R. Soc. A {\bf 209} (1951) 248.

\bibitem{Mashhoon}
B.\,Mashhoon,
Entropy {\bf 23} (2021) 445.

\bibitem{Gorbatsievich}
A.K.\,Gorbatsievich,
Izv. Akad. Nauk BSSR {\bf 2} (1979) 62; Acta Phys. Pol. B {\bf 17} (1986) 111.

\bibitem{Audretsch}
J.\,Audretsch,
J. Phys. A: Math. Gen. {\bf 14} (1981) 411.

\bibitem{Cianfrani&Montani}
F.\,Cianfrani, G.\,Montani,
Int. J. Mod. Phys. A {\bf 23} (2008) 1274; EPL {\bf 84} (2008) 30008.

\bibitem{Obukhov&Silenko&Teryaev}
Y.N.\,Obukhov, A.J.\,Silenko, O.V.\,Teryaev,
Phys. Rev. D {\bf 88} (2013) 084014.

\bibitem{Bjerrum-Bohr&etal-2015-1}
N.E.J.\,Bjerrum-Bohr \emph{et al.},
Phys. Rev. Lett. {\bf 114} (2015) 061301.

\bibitem{Donoghue&El-Menoufi}
J.F.\,Donoghue, B.K.\,El-Menoufi,
J. High Energy Phys. {\bf 05} (2015) 118.

\bibitem{Bjerrum-Bohr&etal-2015-2}
N.E.J.\,Bjerrum-Bohr \emph{et al.},
Int. J. Mod. Phys. D {\bf 24} (2015) 1544013.

\bibitem{Pirani}
F.A.E.\,Pirani,
Acta Phys. Pol. {\bf 15} (1956) 389.

\bibitem{Rumpf}
H.\,Rumpf,
in \emph{Cosmology and Gravitation}, edited by
P.G.\,Bergmann and V.D.\,Sabbata (Plenum Press, 1979).

\bibitem{Manasse&Misner}
F.K.\,Manasse, C.W.\,Misner,
J. Math. Phys. {\bf 4} (1963) 735.

\bibitem{Audretsch-2}
J.\,Audretsch,
Phys. Rev. D {\bf 24} (1981) 1470.

\bibitem{Czrnecka&Czarnecki}
A.P.\,Czarnecka, A.\,Czarnecki,
Am. J. Phys. {\bf 89} (2021) 634.

\bibitem{Laemmerzahl}
C.\,L\"{a}mmerzahl,
Gen. Rel. Grav. {\bf 28} (1996) 1043.

\bibitem{Kulin&etal}
G.V.\,Kulin \emph{et al.},
Nucl. Instrum. Meth. A {\bf 792} (2015) 38.

\bibitem{Hehl&etal}
F.W.\,Hehl, J.\,Lemke, E.W.\,Mielke,
in \emph{Geometry and Theoretical Physics}, edited by
J.\,Debrus and A.C.\,Hirshfeld (Springer-Verlag, 1991).

\bibitem{Mashhoon-1988}
B.\,Mashhoon, Phys. Rev. Lett. {\bf 61} (1988) 2639.

\bibitem{Haag}
R.\,Haag,
\emph{Local Quantum Physics. Fields, Particles, Algebras} (Springer-Verlag, 1996).

\bibitem{DeWitt-1965}
B.S.\,DeWitt,
\emph{Dynamical Theory of Groups and Fields} (Gordon and Breach, 1965).

\bibitem{Overstreet&etal}
C.\,Overstreet \emph{et al.}, Science {\bf 375} (2022) 226.

\bibitem{Gisin&Thew}
N.\,Gisin, R.\,Thew, 
Nature Photon. {\bf 1} (2007) 165.

\bibitem{Rideout&etal}
D.\,Rideout \emph{et al.}, 
Class. Quantum Grav. {\bf 29} (2012) 224011.

\bibitem{Vallone&etal}
G.\,Vallone \emph{et al.},
Phys. Rev. Lett. {\bf 115} (2015) 040502.

\bibitem{Yin&etal}
J.\,Yin \emph{et al.},
Science {\bf 356} (2017) 1140.

\bibitem{Zeldovich&Novikov}
Ya.B.\,Zeldovich, I.D.\,Novikov,
Sov. Astr. {\bf 10} (1967) 602.

\bibitem{Hawking}
S.W.\,Hawking,
Mon. Not. R. astr. Soc. {\bf 152} (1971) 75.

\bibitem{Oppenheimer&Snyder}
J.R.\,Oppenheimer, H.\,Snyder,
Phys. Rev. {\bf 56} (1939) 455.

\bibitem{Casola&etal}
E.\,Di Casola, S.\,Liberati, S.\,Sonego,
Am. J. Phys. {\bf 83} (2015) 39.

\bibitem{Dirac}
P.A.M.\,Dirac,
Can. J. Phys. {\bf 33} (1955) 650.

\bibitem{Pauli}
W.E.\,Pauli,
in \emph{Exclusion principle and quantum mechanics} (Nobel lecture, 1946).

\bibitem{Zeldovich}
Ya.B.\,Zeldovich,
Sov. Phys. Usp. {\bf 11} (1968) 381.

\bibitem{Weinberg-1989}
S.\,Weinberg,
Rev. Mod. Phys. {\bf 61} (1989) 1.

\end{thebibliography}
\end{document}